\documentclass[aps,prd,twocolumn,nobibnotes,nofootinbib,showpacs]{revtex4-1} 

\usepackage{times}
\usepackage{mathptmx}
\usepackage{amsmath}
\usepackage{amsfonts}
\usepackage{amssymb}
\usepackage{mathrsfs}
\usepackage{graphicx}
\usepackage{color}
\usepackage[colorlinks]{hyperref}

\DeclareSymbolFontAlphabet{\mathrsfs}{rsfs}

\definecolor{CiteColor}{rgb}{0,0.5,0}
\hypersetup{citecolor=CiteColor}
\definecolor{RefColor}{rgb}{0.55,0,0}
\hypersetup{linkcolor=RefColor}

\newcommand{\scri}{\mathrsfs{I}}
\newcommand{\D}{\mathcal{D}}
\newcommand{\V}{\mathcal{V}}
\newcommand{\ud}{\mathrm{d}}
\newcommand{\ui}{\mathrm{i}}
\newcommand{\beq}{\begin{equation}}
\newcommand{\eeq}{\end{equation}}
\newcommand{\bsubeq}{\begin{subequations}}
\newcommand{\esubeq}{\end{subequations}}

\begin{document}

\title{Thermodynamics of a Black Hole with Moon}

\author{Samuel E. Gralla}

\author{Alexandre Le Tiec}

\affiliation{Maryland Center for Fundamental Physics \& Joint Space-Science Institute, Department of Physics, University of Maryland, College Park, MD 20742, USA}

\begin{abstract}
For a Kerr black hole perturbed by a particle on the ``corotating'' circular orbit (angular velocity equal to that of the unperturbed event horizon), the black hole remains in equilibrium in the sense that the perturbed event horizon is a Killing horizon of the helical Killing field. The associated surface gravity is constant over the horizon and should correspond to the physical Hawking temperature. We calculate the perturbation in surface gravity/temperature, finding it negative: the moon has a cooling effect on the black hole. We also compute the change in horizon angular frequency, which is positive, and the change in surface area/entropy, which vanishes.
\end{abstract}


\maketitle

\section{Introduction}

Hawking's spectacular realization \cite{Ha.75} that black holes radiate at a temperature given by their horizon surface gravity was a watershed moment in theoretical physics, endowing the classical laws of black hole mechanics \cite{Ha.72,Be2.73,Ba.al.73} with a genuine thermodynamical status. With the result established for stationary, isolated black holes, a natural step forward would be to develop a fully general theory of radiating black holes. However, even the identification of a classical notion of surface gravity for a general black hole is problematic, with different proposals yielding different results, even in spherical symmetry \cite{NiYo.08}. The main difficulty is the lack of a horizon Killing field, whose existence is crucial to the standard notion.

In this paper we study a \textit{dynamical} black hole spacetime for which the surface gravity can nevertheless be unambiguously defined: a Kerr black hole perturbed by a corotating moon. This system is neither stationary nor axisymmetric, as it contains an orbiting particle and its associated gravitational radiation. However, the circular-orbit assumption leads to a \textit{helical} symmetry of the spacetime, and---crucially---the corotation ensures that the orbits of this symmetry coincide with those of the null geodesic generators of the horizon. Thus the event horizon is a Killing horizon, and the surface  gravity $\kappa$ may be defined in the usual way. Standard calculations then show that $\kappa$ is constant on the horizon (a ``zeroth law''), and we argue that the Hawking temperature of the perturbed black hole must still be given by $\hbar \kappa / 2\pi$.

We are able to compute the perturbation in surface gravity in closed form, Eq.~\eqref{Dkappa} below. The perturbation is negative (see Fig.~\ref{fig:Dkappa}), showing that the tidally-induced deformation of the black hole horizon has a \textit{cooling} effect. Recalling that the surface gravity of a Kerr hole decreases with increasing spin, a heuristic picture emerges whereby deformation (whether rotationally, tidally, or otherwise induced) yields a decrease in temperature. We also obtain formulas for the changes in horizon area and angular velocity, Eqs.~\eqref{DA} and \eqref{DomegaH} below. While the area perturbation vanishes, the rotation frequency perturbation is positive (see Fig.~\ref{fig:DomegaH}). Our derivation of the above results features two first-law-type expressions that relate nearby black-hole-moon spacetimes, Eqs.~\eqref{1st_law} and \eqref{de} below.

Some of the inspiration for our work comes from \cite{Fr.al.02}, where zeroth and first laws were established for a class of helically symmetric spacetimes satisfying certain assumptions. In general relativity, however, such spacetimes cannot be asymptotically flat \cite{Fr.al.02,GiSt.84,De.89,Kl.04}. Heuristically, incoming radiation is required to balance the emitted radiation, yielding standing waves whose energy content would not decay fast enough. The lack of smooth asymptotics removes the preferred normalization of the Killing field, and, as emphasized by the authors of \cite{Fr.al.02}, the numerical value of the surface gravity is undetermined.

In this work we avoid those difficulties by using the approximation of a small perturbing moon. To linear order in the size and mass of the moon, gravitational radiation-reaction is absent and incoming radiation is not needed to preserve the helical symmetry. (Physically, our approximation is valid over timescales where backreaction is a small effect.) The spacetime is asymptotically flat at null infinity, where the Killing field may be normalized relative to the time direction of a stationary observer.  While our corotating setup is not generic, it can be realized in nature,\footnote{Stellar-mass compact objects orbiting supermassive black holes are a main target of the planned, space-based gravitational-wave detector eLISA \cite{Am.al.13}.} and our results provide an example of a realistic, \textit{interacting} black hole whose Hawking temperature is well-defined.

The remainder of this paper is organized as follows: Sec.~\ref{sec:spacetime} lays out our perturbative setup. The zeroth and first laws are established in Secs.~\ref{sec:0th_law} and \ref{sec:1st_law}. We compute the perturbations in horizon surface area, angular frequency, and surface gravity in Sec.~\ref{sec:pert}, and discuss a ``physical process'' version of our main results in Sec.~\ref{sec:physical}. Our conventions are those of Ref.~\cite{Wal}. In particular, the metric signature is $(-+++)$ and we use ``geometrized units,'' where $G = c = 1$. Latin indices $a,b,\cdots$ are abstract, while Greek indices $\mu,\nu,\cdots$ are used for coordinate components in a particular coordinate system.

\section{Spacetime of a black hole with moon}\label{sec:spacetime}

\begin{figure}
	\includegraphics[width=0.81\linewidth]{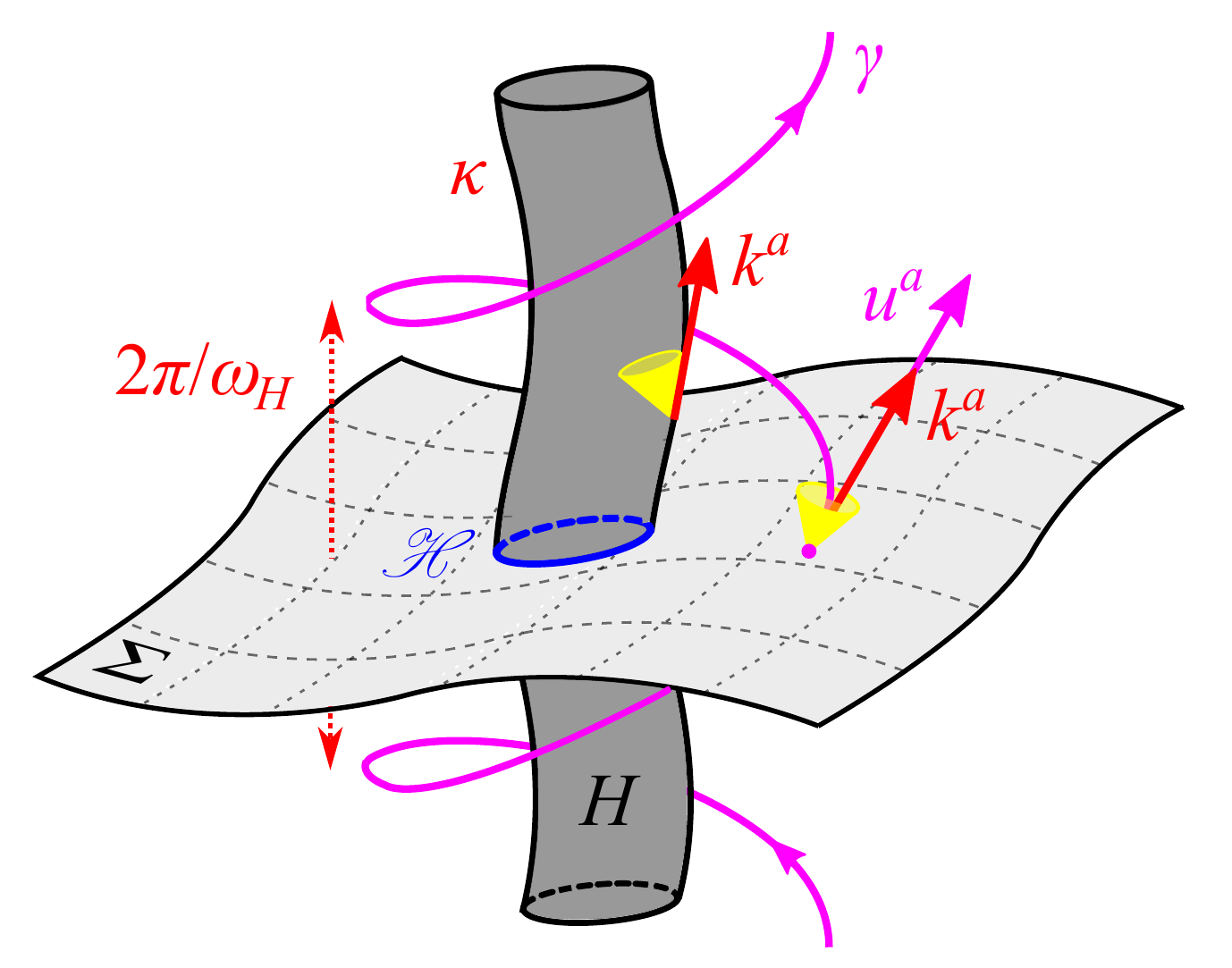}
	\caption{Spacetime diagram depicting a black hole tidally perturbed by a corotating moon. (One spatial dimension is not shown.)} 
	\label{fig:binary}
\end{figure}

We consider a binary system consisting of a black hole orbited by a much smaller moon (see Fig.~\ref{fig:binary}). To obtain an approximate description of this physical system we imagine attaching a one-parameter family of spacetimes $g_{ab}(\lambda)$ to this solution, where the size and mass of the moon are taken to zero with the parameter $\lambda$. The true spacetime is then approximated by a Taylor expansion,
\beq
	g_{ab}(\lambda) = \bar{g}_{ab} + \lambda \, \D g_{ab} + \mathcal{O}(\lambda^2) \, ,
\eeq
where the moon is not present at lowest order ($\lambda = 0$). Here and throughout we use an overbar to denote background quantities and a curly D to denote perturbations, \textit{e.g.}, $\bar{g} \equiv g|_{\lambda=0}$ and $\D g  \equiv \partial_\lambda g|_{\lambda=0}$.

In Ref.~\cite{GrWa.08}, it was shown that for a general body suitably scaled to zero size and mass, the perturbation $\D g_{ab}$ obeys the linearized Einstein equation with point-particle source,
\beq\label{LEE}
	\D G_{ab} = 8 \pi \, \D T_{ab} = 8 \pi \, m \int_\gamma \ud \tau \, \delta_4(x,y) \, u_a u_b \, ,
\eeq
where the curve $\gamma$ is a timelike geodesic of the background. (Here $y(\tau)$ parametrizes the curve $\gamma$ with respect to the background proper time $\tau$,  $u^\mu = \ud y^\mu / \ud \tau$ is the four-velocity, and $\delta_4(x,y)$ is the invariant Dirac distribution in four-dimensional spacetime.) We emphasize that the use of a point particle is \textit{not} a statement about the composition of our body, but rather a consequence of considering an arbitrary body in the limit of small size. The constant parameter $m$ has the interpretation of the ADM mass of the moon as measured in its near-zone \cite{GrWa.08}.

We choose our background metric $\bar{g}_{ab}$ to be the Kerr geometry of mass $\bar{M}$ and angular momentum $\bar{J}$. The black hole horizon has surface area $\bar{A} = 8\pi \, \bar{M}^2 (1+\Delta)$, angular velocity $2 \bar{M} \bar{\omega}_H = \chi/(1+\Delta)$, and surface gravity $2\bar{M} \bar{\kappa} = \Delta/(1+\Delta)$, where $\chi \equiv \bar{J}/\bar{M}^2$ and $\Delta \equiv (1 - \chi^2)^{1/2}.$ We denote the timelike Killing field (normalized to $-1$ at infinity) by $t^a$ and the axial Killing field (with integral curves of parameter length $2\pi$) by $\phi^a$.  We take the geodesic $\gamma$ to be the (unique) equatorial, circular orbit of azimuthal frequency $\bar{\omega}_H$. From the analysis of \cite{Ba.al.72} one may check that this orbit exists and is timelike for all values of $0 < \chi < 1$. However, the orbit is stable only for $\chi<\chi_\text{max} \simeq 0.36$. We denote the conserved orbital quantities associated with $t^a$ and $\phi^a$ by
\bsubeq\label{e_j}
	\begin{align}
		e & = - m \, t^a u_a = m \, \frac{1 - 2 v^2 + \chi v^3}{(1 - 3 v^2 + 2 \chi v^3)^{1/2}} \, , \\
		j & = m \, \phi^a u_a = m \, \frac{\bar{M}}{v} \, \frac{1 - 2 \chi v^3 + \chi^2 v^4}{(1 - 3 v^2 + 2 \chi v^3)^{1/2}} \, ,
	\end{align}
\esubeq
where $v^3 \equiv \bar{M} \bar{\omega}_H / (1 - \chi \bar{M} \bar{\omega}_H)$, and we will refer to $e$ and $j$ as the energy and angular momentum of the particle, respectively. We also introduce the conserved orbital quantity associated with the Killing field $t^a+\bar{\omega}_H \phi^a$, sometimes referred to as the ``redshift observable'' \cite{De.08,Le.al.12},
\beq\label{z}
	z = m^{-1} \left( e - \bar{\omega}_H j \right) = {(1 + \chi v^3)}^{-1} {(1 - 3 v^2 + 2 \chi v^3)}^{1/2} \, .
\eeq

A strategy for constructing the physically-relevant solution of Eq.~\eqref{LEE} is given in Ref.~\cite{Ke.al2.10}. One first solves the Teukolsky equation for the perturbed Weyl scalar $\D \psi_0$, making a choice of no incoming radiation. A ``radiative'' metric perturbation $\D g_{ab}^\text{rad}$ is then constructed from $\D \psi_0$ using the procedure developed in Refs.~\cite{CoKe.74,Ch.75,Wa.78,KeCo.79,St.79}. This perturbation is regular on the future horizon and asymptotically flat at future null infinity. In a suitable gauge, the Boyer-Lindquist coordinate components $\D g^\text{rad}_{\mu \nu}$ depend on Boyer-Lindquist $\phi$ and $t$ only in the combination $\phi-\bar{\omega}_H t$, showing that $\D g_{ab}^\text{rad}$ preserves the helical symmetry of the source $\D T_{ab}$.

However, this perturbation does not satisfy \eqref{LEE} on its own, requiring an additional ``nonradiative'' piece $\D g_{ab}^{\text{NR}}$ to cancel a remaining piece of the source. Since $\D g_{ab}^\text{NR}$ must not change $\D \psi_0$, it must agree with linearized Kerr away from the point particle \cite{Wa.78,Ke.al2.10}, and may be matched at the source to ensure that the linearized Einstein equation is satisfied. (In particular, the non-radiative piece will preserve the helical symmetry of the source, so that the entire perturbation is helically symmetric.) The only remaining freedom is that of a global linearized Kerr perturbation, which is fixed by the requirement that the perturbations in the Bondi\footnote{Since the source has been radiating for all time, gravitational waves reach spatial infinity and the usual falloff conditions required for ADM quantities are not satisfied.} mass $M$ and angular momentum $J$ are given by $\D M = e$ and $\D J = j$. (Alternatively, one can demand that the nonradiative piece gives no contribution to the local Komar mass and angular momentum of the black hole \cite{Ke.al2.10}, or equivalently that $\D g_{ab}^{\text{NR}}$ is pure gauge inside the orbit.) This choice ensures that the perturbation is ``entirely due to the particle,'' with no spurious extra perturbation towards a nearby Kerr solution.

The application of the above procedure requires numerical integration as well as some analytic work on the non-radiative piece that has not yet been carried out. Fortunately, we will not need full details of the perturbed spacetime $\bar{g}_{ab} + \lambda \D g_{ab}$ to establish our results, and can rely instead on the general properties established above: helical symmetry, no incoming radiation, regularity on the future horizon, and asymptotic flatness at future null infinity. We will show that these properties imply the vanishing of the expansion and shear of the geodesic generators of the perturbed event horizon. By rigidity arguments \cite{Ha.72,HaEl,Ch.96,Ch.97,Fr.al.99,Al.al.10} this should in turn imply that the perturbed horizon is a Killing horizon.\footnote{Any particular version of the rigidity theorem has a set of technical assumptions, such as analyticity and/or a bifurcate horizon structure, in addition to the essential physical requirement of vanishing expansion and shear. We have not checked whether our metric perturbation satisfies the specific assumptions of a particular version of the theorem.} The perturbed horizon then has a well-defined surface gravity and angular frequency, and we are able to compute these quantities (along with the perturbed surface area) without further assumptions about the spacetime.

\section{Zeroth law}\label{sec:0th_law}

\subsection{Killing Horizon and Surface Gravity}

Since our perturbed spacetime $g_{ab} + \lambda \D g_{ab}$ becomes singular along the worldline $\gamma$, we cannot directly define the (future) event horizon $H$ as the boundary of the past of future null infinity. However, we may still employ this definition within the one-parameter family $g_{ab}(\lambda)$ at any finite $\lambda$, and it is clear on physical grounds that $H$ will be a smooth function of $\lambda$ at $\lambda=0$, ensuring that the perturbed event horizon is well-defined.\footnote{If the small body is also a black hole, the event horizon will be disjoint for all $\lambda>0$. As $\lambda \rightarrow 0$ and the small hole disappears, however, it seems clear that the large horizon will behave smoothly, and it is the perturbation of this component of the horizon that we study.} If we can establish that the expansion and shear of the horizon vanish to $\mathcal{O}(\lambda)$, then rigidity arguments should ensure that the event horizon is a Killing horizon. We may then appeal to \cite{Ba.al.73}, who established the constancy of the surface gravity $\kappa^2 = \frac{1}{2} \nabla^a k^b \nabla_b k_a$ of any Killing horizon generated by a Killing field $k^a$, in any spacetime satisfying the dominant energy condition (at least locally). In particular, this establishes the constancy of $\D \kappa$ for our locally vacuum spacetime.

It remains to show the vanishing of the perturbed expansion and shear. To do so, we introduce (at any $\lambda \geqslant 0$) a Newman-Penrose (NP) \cite{NePe.62} tetrad\footnote{The real null vectors $\ell^a$ and $n^a$ satisfy $\ell^a n_a=-1$, while the complex null vector $m^a$ satisfies $m^a m_a^* = 1$. All other inner products vanish.} $\{\ell^a, n^a, m^a, m^{*a}\}$ such that $\ell^a$ is tangent to the null geodesic generators of the horizon, while $m^a$ and $m^{*a}$ are parallel-transported along those generators. (A star denotes complex conjugation.) The expansion and shear will vanish if and only if the spin coefficients $\rho = - m^a m^{*b} \nabla_b \ell_a$ and $\sigma = - m^a m^b \nabla_b \ell_a$ are vanishing on $H$. For our tetrad on the horizon, the NP equations for $\rho$ and $\sigma$ become
\bsubeq\label{animals}
	\begin{align}
		\ell^a \nabla_a \rho & = \rho^2 + \sigma \sigma^* + 2 \epsilon \rho \, , \label{horse} \\
		\ell^a \nabla_a \sigma & = 2 \rho \sigma + 2 \epsilon \sigma + \psi_0 \, , \label{cow}
	\end{align}
\esubeq
where $2\epsilon = - n^a \ell^b \nabla_b \ell_a$ and $\psi_0 = \!\: C_{abcd} \ell^a m^b \ell^c m^d$.\footnote{For our choice of tetrad on $H$, the spin coefficient $\kappa$ vanishes (generators are geodesic), $\rho$ is real (generators are hypersurface orthogonal), and $\epsilon$ is real ($m^a$ and $m^{*a}$ are parallel-transported).} Equations \eqref{animals} hold at finite $\lambda$ on $H$. We now normalize the tetrad at $\lambda=0$ such that $\ell^a = t^a + \bar{\omega}_H \phi^a$ on the unperturbed horizon. Then, $2 \bar{\epsilon}$ coincides with the surface gravity $\bar{\kappa}$ of the Kerr spacetime. Furthermore, we have that $\rho$, $\sigma$, and $\psi_0$ all vanish when $\lambda=0$, so that the perturbation of Eqs.~\eqref{animals} gives
\bsubeq\label{deltas}
	\begin{align}
		(t^a + \bar{\omega}_H \phi^a) \nabla_a \D \rho & = \bar{\kappa} \, \D \rho \, , \label{drho} \\
		(t^a + \bar{\omega}_H \phi^a) \nabla_a \D \sigma & = \bar{\kappa} \, \D \sigma + \D \psi_0 \, . \label{dsigma}
	\end{align}
\esubeq
However, the left-hand sides of \eqref{deltas} vanish by the helical symmetry of the perturbed spacetime. Since we consider only the non-extremal case $\chi<1$ we have $\bar{\kappa} \neq 0$, and it follows that
\beq\label{Dsigma}
	\D \rho = 0 \quad \text{and} \quad \D \sigma = - \bar{\kappa}^{-1} \D \psi_0 \, .
\eeq
We now take advantage of the Teukolsky equation to compute $\D \psi_0$ on the horizon. Equations (4.43), (4.40), and (4.42) of Ref.~\cite{TePr.74} show that  each mode of $\D \psi_0$, say ${(\D \psi_0)}_{\ell m \omega}$, is given near the horizon ($r_* \rightarrow -\infty$) by
\begin{align}\label{porcupine}
	{(\D \psi_0)}_{\ell m \omega} & \sim A_{\ell m \omega} \,\, \ui \tilde{k} \, (\tilde{k}^2+\bar{\kappa}^2)(-\ui\tilde{k}+2\bar{\kappa}) \nonumber \\ & \qquad \qquad \times { }_2S_{\ell m \omega}(\theta,\phi) \, e^{-\ui(\tilde{k} r_* + \omega t)} , 
\end{align}
where $(t,r,\theta,\phi)$ are Boyer-Lindquist coordinates, with $r_*$ the tortoise coordinate, ${ }_2S_{\ell m \omega}(\theta,\phi)$ are spin-weighted spheroidal harmonics, $\tilde{k} \equiv \omega - m \bar{\omega}_H$, and the amplitudes $A_{\ell m \omega}$ are determined by solving the Teukolsky equation. (Note that our $\D \psi_0$ corresponds to their $\psi_0^{\,\,\text{HH}}$.) However, when a circular orbit of frequency $\bar{\omega}_H$ is assumed, and no incoming radiation is chosen, the full field $\D \psi_0$ is given by a sum over modes with $\omega = m \bar{\omega}_H$, \textit{i.e.}, we have $\tilde{k}=0$. Then Eq.~\eqref{porcupine} gives $\D \psi_0 = 0$ on $H$, and from \eqref{Dsigma} we conclude that $\D \sigma=0$.

\subsection{Horizon Killing Field and Angular Velocity}

The above argument establishes the existence of a horizon Killing field to $\mathcal{O}(\lambda)$, \textit{i.e.}, of a vector field $k^a(\lambda)$ satisfying $\mathcal{L}_k g_{ab}=\mathcal{O}(\lambda^2)$ (at least in a neighborhood of the horizon) and normal to $H$. In addition to the helical Killing field of the metric perturbation (proportional to $t^a+\bar{\omega}_H \phi^a$ in a gauge, such as that of Ref.~\cite{Ke.al2.10}, where the metric components $\D g_{\mu\nu}$ are asymptotically vanishing), our perturbed spacetime also possesses the trivial Killing fields $\lambda t^a$ and $\lambda \phi^a$ inherited from the symmetries of the background. By a choice of normalization we may eliminate the perturbation to $t^a$, and the horizon Killing field can be written as
\beq\label{yak}
	k^a(\lambda) = t^a + \left( \bar{\omega}_H + \lambda \D \omega_H \right) \phi^a + \mathcal{O}(\lambda^2) \, ,
\eeq
where $\D \omega_H$ is a constant. A gauge transformation that changes the value of $\D \omega_H$ would introduce metric components that are not asymptotically Minkowskian. Therefore, if we stipulate that the metric components be asymptotically Minkowskian, and that $k^a(\lambda)$ be normal to the horizon, Eq.~\eqref{yak} defines $\D \omega_H$ as an intrinsic, coordinate-invariant property of the perturbed spacetime. Since $t^a$ and $\phi^a$ represent the time and rotational directions of a distant stationary observer, this constant can be interpreted as the perturbation in the horizon angular velocity. In work on comparable mass-ratio binaries, $\omega_H = \bar{\omega}_H + \lambda \D \omega_H$ is sometimes interpreted as the circular-orbit frequency of the binary system \cite{Go.al.02,Sh.al.04}.

\begin{figure}[t!]
	\includegraphics[width=0.78\linewidth]{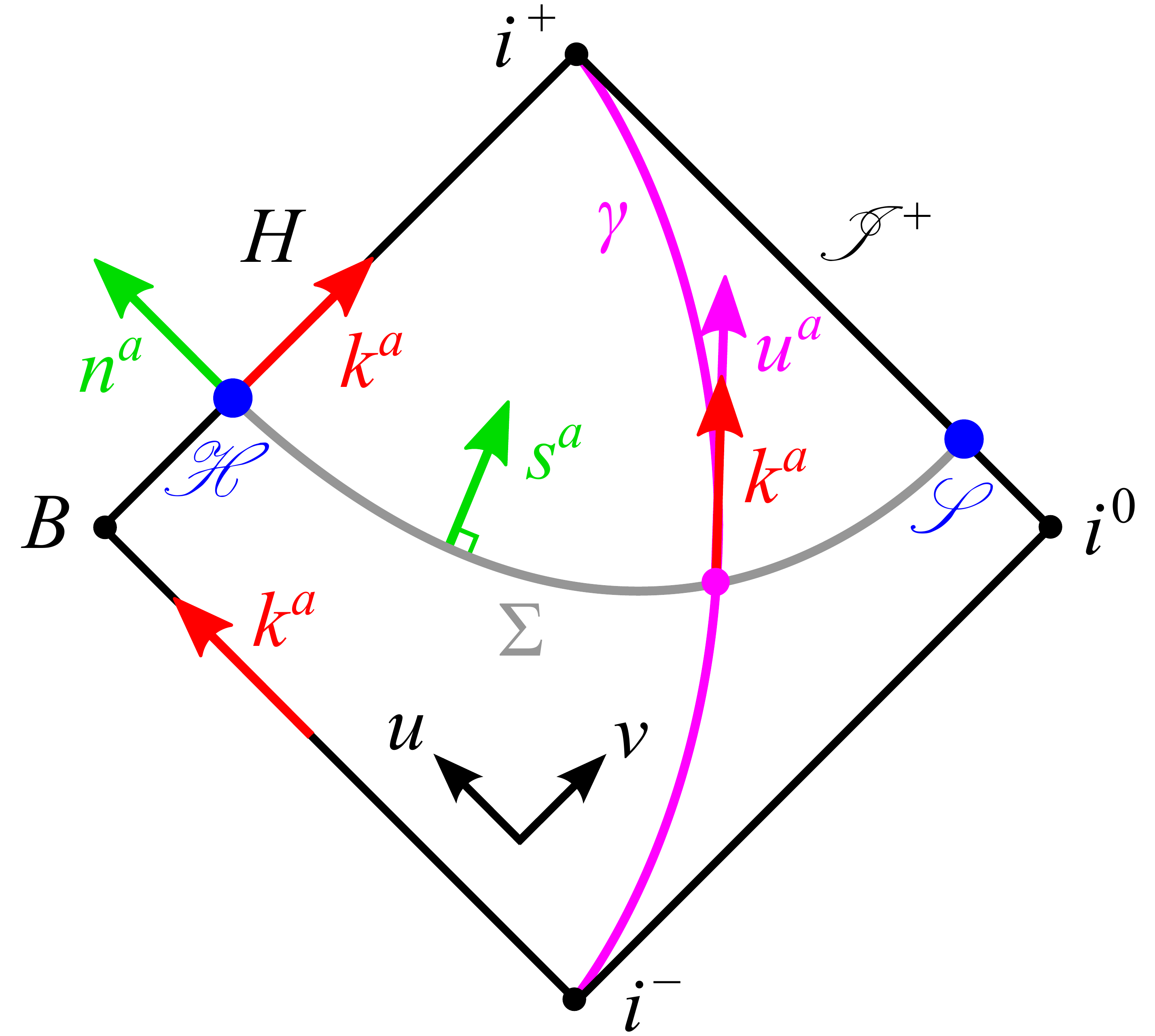}
	\caption{Carter-Penrose diagram showing the causal structure of our perturbative spacetime. (Two spatial dimensions are not shown.)} 
	\label{fig:CP}
\end{figure}

\subsection{Hawking Temperature}

We now argue that the surface gravity $\kappa = \bar{\kappa} + \lambda \D \kappa$ of our tidally perturbed black hole still coincides with the physical Hawking temperature $T_\text{H}$. Our main point is that all of the essential properties underlying the semi-classical calculation for Kerr are preserved in our spacetime. In particular, we have a horizon Killing field $k^a = t^a + \omega_H \phi^a$, infinitesimally related to that of Kerr, which is normalized so that $k^a t_a=-1$ at infinity. The main new complication with respect to the Kerr case is that $t^a$ and $\phi^a$ are not separate symmetries of our perturbed spacetime. However, $t^a$ remains an \textit{asymptotic} time translation symmetry, which may be used to define positive and negative frequency modes with respect to a distant stationary detector, and the usual wavepacket scattering experiment may still be posed. We expect that, just as in the Kerr case, there will be a mixing of positive and negative frequency modes controlled by the surface gravity $\kappa$ associated with the Killing field $k^a$, leading to a particle flux through the distant detector with characteristic temperature
\beq
	T_\text{H} = \frac{\hbar}{2\pi} \, \kappa \, .
\eeq
The lack of separate stationarity and axisymmetry will make this flux time and angle-dependent (though it must respect the helical symmetry), which a detailed calculation would presumably characterize in terms of a suitable ``greybody factor'' modifying the Planck spectrum.  

It should be pointed out, however, that there will likely be additional particle creation effects not associated directly with the black hole. Particle creation is a generic feature of non-stationary spacetimes, and such effects would persist if, \textit{e.g.}, we considered a rotating star instead of a black hole. However, in our spacetime these effects should scale with the orbital frequency rather than the surface gravity and should be readily distinguishable from the Hawking radiation of the black hole. In this context we would still regard $\kappa$ as the temperature of the black hole.

\section{First Laws}\label{sec:1st_law}

\subsection{First Law for a Black Hole with Moon}

We shall now establish a first law of mechanics that relates two such nearby ``black hole with corotating moon'' solutions. We will therefore consider a \textit{two}-parameter family of spacetimes $g_{ab}(\lambda,\epsilon)$, where for each $\epsilon$, $g_{ab}(\lambda,\cdot)$ is a one-parameter family of the type discussed in section \ref{sec:spacetime}. We will denote by $\delta g \equiv \partial_\epsilon g \vert_{\epsilon=0}$ the variation of the metric under small changes in the properties of a black-hole-moon spacetime at fixed $\lambda$, and similarly for all other quantities. Any quantity without a $\delta$ in front will refer to the $\epsilon=0$ spacetime, in the usual manner of variational calculations. We will work consistently to $\mathcal{O}(\lambda)$, dropping higher-order terms.

Iyer and Wald \cite{IyWa.94} gave a general derivation of the first law for arbitrary vacuum perturbations of a stationary black hole that are asymptotically flat at spatial infinity. Here we follow their general strategy, while making appropriate modifications for our non-vacuum perturbations of a non-stationary black hole spacetime that are asymptotically flat at null infinity. Let $\Sigma$ be an arbitrary spacelike slice transverse to a Killing field $k^a$ of the background ($\epsilon=0$), with boundary $\partial \Sigma$. Then calculations similar to those performed in Refs.~\cite{IyWa.94,GaWa.01} yield the identity
\begin{align}\label{snail}
	\frac{1}{16 \pi} \int_{\partial\Sigma} \bigl( \delta Q_{ab} - k^c \Theta_{abc} & \bigr) = \delta \int_\Sigma \epsilon_{abcd} \, T^{de} k_e \nonumber \\ \, &- \frac{1}{2} \int_\Sigma \epsilon_{abcd} \, k^d \, T^{ef} \delta g_{ef} \, ,
\end{align}
where $Q_{ab} = - \epsilon_{abcd} \nabla^c k^d$ is the Noether charge two-form associated with $k^a$, and $\Theta_{abc} = \epsilon_{abcd} \, g^{de} g^{fh} (\nabla_e \delta g_{fh} - \nabla_f \delta g_{eh})$ the symplectic potential three-form of general relativity, with $\epsilon_{abcd}$ the natural volume element associated with $g_{ab}$. In equation \eqref{snail} $k^a$ is a fixed vector field; hence $\delta k^a=0$. Hereafter, we further specify $\Sigma$ such that its inner and outer boundaries correspond to a cross-section $\mathcal{H} = H \cap \Sigma$ of the future event horizon and a two-sphere $\mathcal{S} = \scri^+ \! \cap \Sigma$ at future null infinity (see Fig.~\ref{fig:CP}); hence $\int_{\partial\Sigma} = \oint_\mathcal{S} - \oint_\mathcal{H}$. We also require that the future-directed, unit normal to $\Sigma$, say $s_a$, coincides with $u_a$ at the point $\gamma \cap \Sigma$. Finally, we take $k^a$ to be the horizon Killing field of the background ($\epsilon=0$), $k^a = t^a + \omega_H \phi^a$.

According to the general analysis of Ref.~\cite{WaZo.00}, the surface integral over $\mathcal{S}$ yields the perturbation in the Bondi quantity associated with the asymptotic symmetry $t^a + \omega_H \phi^a$, \textit{i.e.},
\beq\label{pif}
	\frac{1}{16 \pi} \oint_\mathcal{S} \left( \delta Q_{ab} - k^c \Theta_{abc} \right) = \delta M - \omega_H \delta J \, .
\eeq
This result follows from Eq.~(81) of Ref.~\cite{WaZo.00}, where the term involving $N_{ab} \tau^{ab}$ is $\mathcal{O}(\lambda^2)$ and does not contribute. To see this, note first that the Bondi news tensor $N_{ab}$ is $\mathcal{O}(\lambda)$ because it vanishes for stationary spacetimes.  We may therefore compute $\tau^{ab}$ (defined in Eq.~(50) of \cite{WaZo.00}) having set $\lambda=0$. In this case the variation $\delta$ corresponds to a change of parameters of the Kerr spacetime, and it is easily checked that $\tau^{ab}=0$ for such a variation.

Next, we turn to the surface integral over the cross-section $\mathcal{H}$ of the unperturbed ($\epsilon = 0$) horizon $H$. We make use of the element $n^a$ of the NP tetrad introduced in section \ref{sec:0th_law}. Because $k^a = \ell^a$ and $n^a$ are both normal to $\mathcal{H}$, and satisfy $k^a n_a = -1$ there, the metric volume element can be written as $\epsilon_{abcd} = 12 k_{[a} n_b \tilde{\epsilon}_{cd]}$, where $\tilde{\epsilon}_{ab} = \epsilon_{abcd} n^c k^d$ is the area element on $\mathcal{H}$. Calculations similar to those of \cite{Ba.al.73,Fr.al.02} then yield $Q_{ab} = 2 \kappa \, \tilde{\epsilon}_{ab}$ and $k^c \Theta_{abc} = 2 \delta \kappa \: \tilde{\epsilon}_{ab}$ on $\mathcal{H}$. Finally, $\kappa$ and $\delta \kappa$ being constant on the horizon, the surface integral yields
\beq\label{paf}
	\frac{1}{16 \pi} \oint_\mathcal{H} \left( \delta Q_{ab} - k^c \Theta_{abc} \right) = \frac{\kappa}{8\pi} \, \delta A \, ,
\eeq
where $A = \oint_\mathcal{H} \tilde{\epsilon}_{ab}$ is the surface area of the event horizon.

Finally we consider the volume integrals over $\Sigma$, in which the stress-energy tensor $T^{ab} = \lambda \D T^{ab}$ is given by \eqref{LEE} above. [Because $T^{ab}$ is $\mathcal{O}(\lambda)$, only the $\mathcal{O}(\lambda^0)$ parts of the other quantities will contribute; in particular we have $k^a = t^a + \bar{\omega}_H \phi^a + \mathcal{O}(\lambda)$.] Let $\Sigma$ coincide with a surface $T = \text{const}$ for some scalar field $T$; hence $s_a = - N \, \nabla_a T$, where $N$ is the lapse function. Since $\epsilon_{abcd} = 4 \hat{\epsilon}_{[abc} s_{d]}$ over $\Sigma$, where $\hat{\epsilon}_{abc} = - \epsilon_{abcd} s^d$ is the natural volume element on $\Sigma$, and since $\ud \tau = N \, \ud T$ at the point $\gamma \cap \Sigma$ (because $s_a = u_a$ there), the hypersurface integrals can be computed using the defining property of the invariant Dirac distribution $\delta_4$, namely that $\int_{\V} \epsilon_{abcd} \, f(x) \, \delta_4(x,y) = f(y)$ for any smooth test function $f(x)$ and any four-dimensional region $\V \owns y$. Then, using the colinearity of the helical Killing vector and the four-velocity, $k^a = z \, u^a$, as well as the normalization $u^a u_a=-1$, we find
\bsubeq\label{butterfly}
	\begin{align}
		\int_\Sigma \epsilon_{abcd} \, T^{de} k_e &= \lambda m \, z \, , \\
		\int_\Sigma \epsilon_{abcd} \, k^d \, T^{ef} \delta g_{ef} &= - \lambda m \, z \, u^a u^b \delta g_{ab} = 2 \lambda m \, \delta z \, ,
	\end{align}
\esubeq
where the last equality follows from $u^a u^b \delta g_{ab}= -2 u_a \delta u^a$ and $\delta (z \, u^a) = \delta k^a = 0$ at the particle's location.

Collecting the intermediate results \eqref{snail}--\eqref{butterfly}, and making the slight abuse of notation $\lambda m \to m$, we find
\beq\label{1st_law}
	\delta M = \omega_H \, \delta J + \frac{\kappa}{8\pi} \, \delta A + z \, \delta m \, .
\eeq
This variational relation generalizes the well-known first law of black hole mechanics \cite{Ba.al.73,IyWa.94} to any two neighboring non-stationary, non-axisymmetric, non-vacuum black hole spacetimes with corotating moons.  Similar results were previously established for two black holes \cite{Fr.al.02} and for two point particles \cite{Le.al.12,Bl.al.13}.  The right-hand sides of these first laws involve the sums $\sum_i \kappa_i \, \delta A_i / 8 \pi$ and $\sum_i z_i \, \delta m_i$ over the black holes and the point masses, respectively; here we see that
the first law \eqref{1st_law} for a black hole with a point mass involves one term of each. Equation \eqref{1st_law} was previously written down in Ref.~\cite{Le.al.12}, without attempting to give a precise perturbative meaning to the quantities that appear.

For each $\varepsilon$, the perturbative spacetime $g_{ab}(\lambda,\epsilon)$ is entirely characterized [to $\mathcal{O}(\lambda)$] by, \textit{e.g.}, the mass $m$ of the particle, the surface area $A$ of the perturbed horizon, and the Bondi angular momentum $J$. Since the Einstein equation does not contain any privileged mass scale, the Bondi mass $M$ must be a homogeneous function of degree one in $J^{1/2}$, $A^{1/2}$ and $m$. Hence, applying Euler's theorem to the function $M(J^{1/2},A^{1/2},m)$ and using the first law \eqref{1st_law} immediately gives the first integral
\beq\label{Smarr}
	M = 2 \omega_H J + \frac{\kappa A}{4\pi} + m \, z \,,
\eeq
where, as in \eqref{1st_law}, a factor of $\lambda$ in front of $m z$ has been dropped. This result generalizes Smarr's formula $\bar{M} = 2 \bar{\omega}_H \bar{J} + \bar{\kappa} \bar{A} / 4\pi$ for Kerr black holes \cite{Sm.73}, which is recovered here at $\lambda=0$. Alternatively, the formula \eqref{Smarr} can be established by using the standard identity (see, \textit{e.g.}, Refs.~\cite{Wal,Poi})
\beq\label{identity}
	\frac{1}{8\pi} \int_{\partial \Sigma} Q_{ab} = \frac{1}{4\pi} \int_\Sigma \epsilon_{abcd} \, R^{de} k_e \, ,
\eeq
which is valid for any Killing field in any spacetime, by evaluating each term to $\mathcal{O}(\lambda)$ in our perturbed, helically symmetric spacetime. The integral over $\mathcal{S}$ yields $M - 2 \omega_H J$, the integral over $\mathcal{H}$ gives $\kappa A / 4\pi$, and the integral over $\Sigma$ yields $\lambda m \, z$.

\subsection{Particle Hamiltonian First Law}

We shall now derive a second variational relation, to be used in section \ref{sec:pert} below. The geodesic motion of a \textit{test} mass $m$ in the curved spacetime $\bar{g}_{ab}(x;\bar{M},\bar{J})$ of a Kerr black hole of mass $\bar{M}$ and spin $\bar{J}$ can be derived from the Hamiltonian \cite{Ca.68}
\beq\label{H}
	H(y,p;\bar{M},\bar{J}) = \frac{1}{2} \, \bar{g}^{ab}(y;\bar{M},\bar{J}) \, p_a p_b \, ,
\eeq
where $y$ and $p$ are the particle's canonical position and four-momentum, viewed as functions of the affine parameter $\tau / m$. (In Boyer-Lindquist coordinates, $p_\mu = (-e,0,0,j)$ for equatorial circular orbits.) This dynamical system being completely integrable, the motion can be described by using generalized action-angle variables $(q_\alpha,J_\alpha)$, with $\alpha \in \{0,\cdots,3\}$ \cite{Sc.02,HiFl.08}. Varying the Hamiltonian $H(q_\alpha,J_\alpha;\bar{M},\bar{J})$ with respect to its arguments, and using the equations of motion $\dot{q}_\alpha = \Omega_\alpha$ and $\dot{J}_\alpha = 0$, as well as the ``on shell'' constraint $H = - m^2 / 2$, the following relationship can be established:\footnote{Full details will be given elsewhere \cite{Le.13}. Note that Eq.~\eqref{de} is not merely valid for two nearby corotating circular orbits in two nearby Kerr solutions; this relationship can be generalized to any geodesic orbit in Kerr.}
\beq\label{de}
	\delta e = \bar{\omega}_H \, \delta j + z \, \delta m + \frac{z}{m} \left( \frac{\partial H}{\partial \bar{M}} \, \delta \bar{M} + \frac{\partial H}{\partial \bar{J}} \, \delta \bar{J} \right) .
\eeq
Here, the partial derivatives of the Hamiltonian are to be taken while keeping the canonical variables fixed, and are followed by the reduction to the circular orbit with frequency $\bar{\omega}_H$. This yields the following expressions in terms of $v$ and $\chi$:
\bsubeq\label{dH}
	\begin{align}
		\frac{\partial H}{\partial \bar{M}} &= - \frac{m^2}{\bar{M}} \, v^2 \, \frac{1 + 2 \chi v^3 - \chi^2 v^4}{1 - 3 v^2 + 2 \chi v^3} \, , \\
		\frac{\partial H}{\partial \bar{J}}	&= \frac{m^2}{\bar{M}^2} \, v^5 \, \frac{2 - \chi v}{1 - 3 v^2 + 2 \chi v^3} \, .
	\end{align}
\esubeq

As discussed earlier, because the Einstein equation involves no privileged mass scale, the particle's energy $e$ has to be a homogeneous function of degree one in $\bar{M}$, $\bar{J}^{1/2}$, $m$, and $j^{1/2}$. Applying Euler's theorem to the function $e(\bar{M},\bar{J}^{1/2},m,j^{1/2})$ and using the Hamiltonian first law \eqref{de} yields the first integral
\beq\label{e}
	e = 2 \bar{\omega}_H j + m \, z + \frac{z}{m} \left( \bar{M} \, \frac{\partial H}{\partial \bar{M}} + 2 \bar{J} \, \frac{\partial H}{\partial \bar{J}} \right) .
\eeq
Using Eqs.~\eqref{e_j}, \eqref{z} and \eqref{dH}, one can easily check that \eqref{de} and \eqref{e} are indeed satisfied. Note that although the first laws \eqref{1st_law} and \eqref{de} are conceptually different, they share the schematic form $\delta\text{(energy)} = \omega \, \delta\text{(ang. mom.)} + z \, \delta m + \text{(other terms)}$.

\begin{figure}[t!]
	\includegraphics[width=0.91\linewidth]{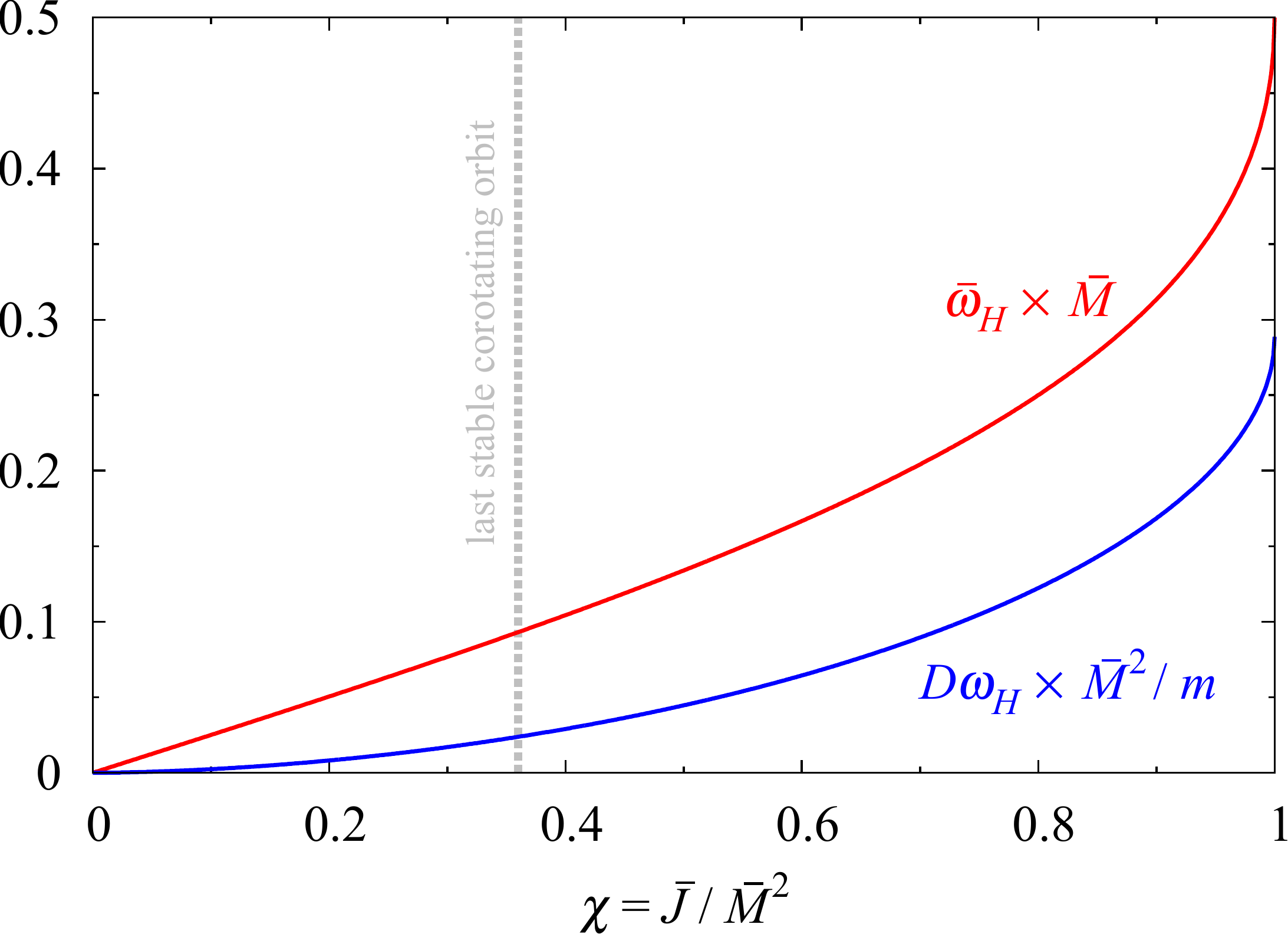}
	\caption{Horizon angular velocity $\bar{\omega}_H$ of a Kerr black hole of mass $\bar{M}$ and spin $\bar{J}$ (red) and the perturbation $\D \omega_H$ induced by a corotating moon of mass $m$ (blue).} 
	\label{fig:DomegaH}
\end{figure}

\section{Perturbations in horizon surface area, angular velocity, and surface gravity}\label{sec:pert}

We will now employ the variational relations and first integrals derived in the previous section to compute the horizon surface area, angular velocity, and surface gravity of a single black-hole-moon spacetime. For the area perturbation, we consider the particular two-parameter family $g_{ab}(\lambda,\epsilon)$ whose background spacetime $g_{ab}(\lambda,0)$ has no moon, $m(\epsilon=0)=0$, while the perturbed spacetime $\delta g_{ab}(\lambda)$ has a moon of mass $m$. This amounts to sending $\delta M \rightarrow \D M$, $\delta J \rightarrow \D J$,  $\delta A \rightarrow \D A$ and $\delta m \rightarrow m$ in Eq.~\eqref{1st_law}, yielding
\beq\label{1st_law_bis}
	\D M = \bar{\omega}_H \, \D J + \frac{\bar{\kappa}}{8\pi} \, \D A + m \, z \, .
\eeq
This equation refers to a single ``black hole with moon'' spacetime. Using $\D M = e$ and $\D J = j$ [see sections \ref{sec:spacetime} and \ref{sec:physical}], as well as $m \, z = e - \bar{\omega}_H j$ [see Eq.~\eqref{z}], we find that the perturbation in horizon surface area vanishes:
\beq\label{DA}
	\D A = 0 \, .
\eeq
Since the entropy of any black hole is proportional to the area $A$ of horizon cross-sections, we see that the orbiting moon does not affect the entropy of the companion hole. (The result \eqref{DA} holds for any cross-section $\mathcal{H}$, a fact that can be seen \textit{a priori} from the helical symmetry of the perturbed spacetime.)

\begin{figure}[t!]
	\includegraphics[width=0.92\linewidth]{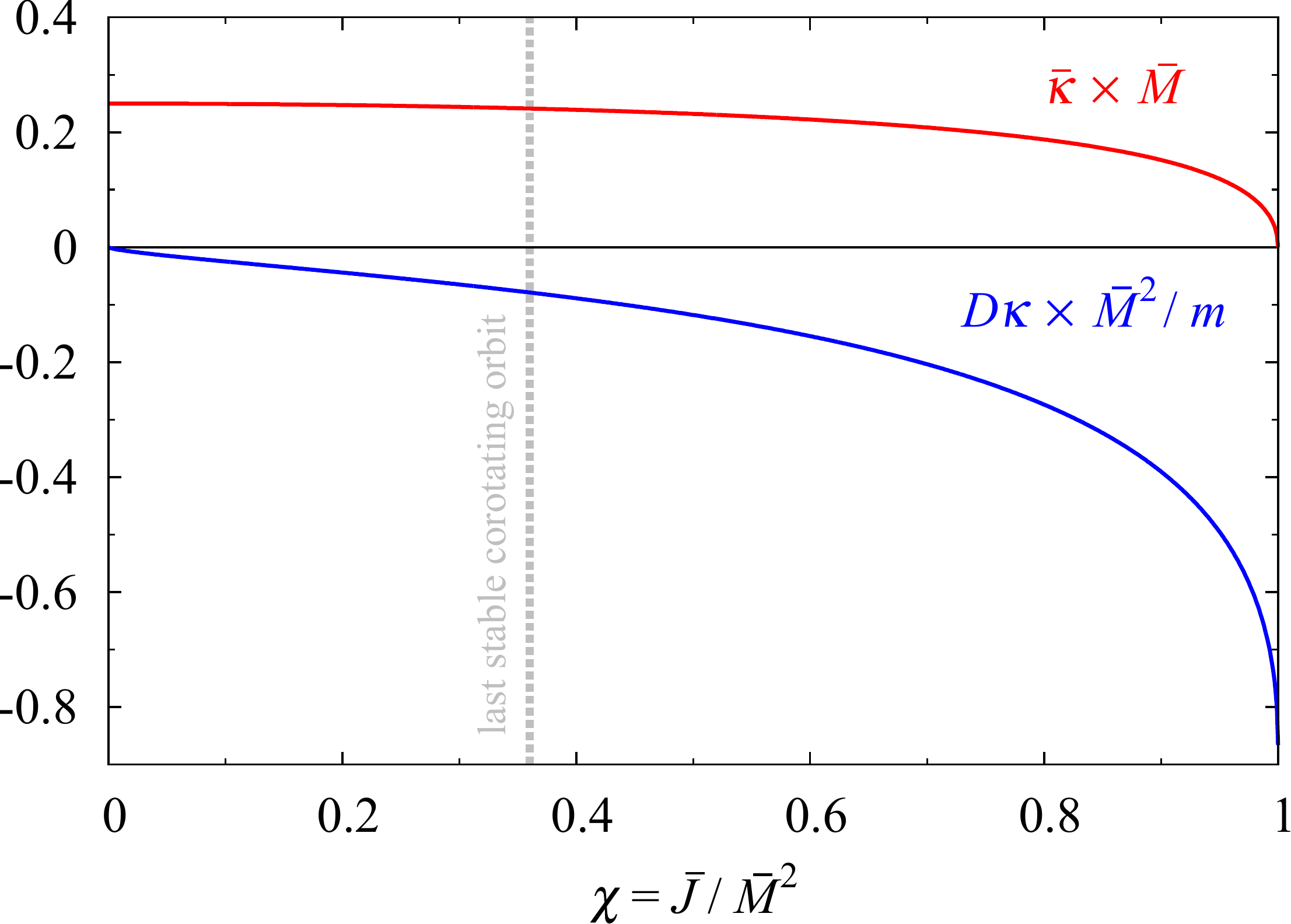}
	\caption{Horizon surface gravity $\bar{\kappa}$ of a Kerr black hole of mass $\bar{M}$ and spin $\bar{J}$ (red) and the perturbation $\D \kappa$ induced by a corotating moon of mass $m$ (blue).} 
	\label{fig:Dkappa}
\end{figure}

For the angular frequency $\D \omega_H$, we subtract Eqs.~\eqref{de} and \eqref{e} from the $\mathcal{O}(\lambda)$ contributions  to the first law \eqref{1st_law} and the first integral relation \eqref{Smarr}, respectively. Using the result \eqref{DA}, we immediately get
\bsubeq
	\begin{align}
		\D \omega_H \, \delta \bar{J} + \frac{\D \kappa}{8\pi} \, \delta \bar{A} &= \frac{z}{m} \left( \frac{\partial H}{\partial \bar{M}} \, \delta \bar{M} + \frac{\partial H}{\partial \bar{J}} \, \delta \bar{J} \right) , \label{ping} \\
		2 \bar{J} \, \D \omega_H + \frac{\bar{A}}{4\pi} \, \D \kappa &= \frac{z}{m} \left( \bar{M} \, \frac{\partial H}{\partial \bar{M}} + 2 \bar{J} \, \frac{\partial H}{\partial \bar{J}} \right) . \label{pong}
	\end{align}
\esubeq
Now we use \eqref{pong} to substitute $\D \kappa$ in favor of $\D \omega_H$ in \eqref{ping}, and replace $\delta \bar{M}$ by $\delta \bar{J}$ and $\delta \bar{A}$ via the ordinary first law, \textit{i.e.,} Eq.~\eqref{1st_law} at $\lambda=0$. Then Eq.~\eqref{ping} reduces to an equation of the form $K \, \delta \ln{(\bar{A}/\bar{J})} = 0$. Since the variations $\delta \bar{A}$ and $\delta \bar{J}$ are independent, we must have $K = 0$, from which we deduce the following change in horizon angular velocity:
\beq\label{DomegaH}
	\D \omega_H = \frac{z}{m} \left( \frac{\partial H}{\partial \bar{J}} + \bar{\omega}_H \, \frac{\partial H}{\partial \bar{M}} \right) .
\eeq
Using \eqref{z} and \eqref{dH}, as well as the expressions for $\bar{\omega}_H$ and $v$ as functions of the Kerr parameter $\chi = \bar{J} / \bar{M}^2$, the perturbation in horizon frequency can be written as $\D \omega_H = (m / \bar{M}^2) \, f(\chi)$, where $f$ is a monotonically increasing function of $\chi$ such that $f(0) = 0$ and $f(1) = 1 / (2\sqrt{3})$ (see Fig.~\ref{fig:DomegaH}). Therefore, $\D \omega_H$ is positive for all spin values: the finite mass $m$ of the moon increases the apparent angular frequency of the black hole.

Finally, we compute the perturbation $\D \kappa$ in surface gravity. Substituting the expression \eqref{DomegaH} for $\D \omega_H$ in \eqref{pong}, and making use of Smarr's formula $\bar{M} = 2 \bar{\omega}_H \bar{J} + \bar{\kappa} \bar{A} / 4\pi$, we obtain the closed-form expression
\beq\label{Dkappa}
	\D \kappa = \bar{\kappa} \,\, \frac{z}{m} \,\, \frac{\partial H}{\partial \bar{M}} \, .
\eeq
Hence the relative change in surface gravity is closely related to the orbital dynamics of the particle. From the expressions for $\bar{\kappa}$, $z$, and $\partial H / \partial \bar{M}$ as functions of the Kerr parameter $\chi$ [see Eqs.~\eqref{z} and \eqref{dH}], the perturbation in surface gravity can be written as $\D \kappa = (m / \bar{M}^2) \, g(\chi)$, where $g$ is a monotonically decreasing function of $\chi$ such that $g(0) = 0$ and $g(1) = - \sqrt{3} / 2$ (see Fig.~\ref{fig:Dkappa}), and is thus negative for all spin values. Note that since the surface gravity $\bar{\kappa}$ of an extremal Kerr black hole vanishes, the perturbation becomes dominant in that limit, signaling the breakdown of the perturbation expansion. This is consistent with the observation made in \cite{Ba.al.73} that nearly-extremal black holes are ``loosely bound,'' in the sense that a small perturbation can raise a large tide. For the last stable corotating circular orbit ($\chi_\text{max} \simeq 0.36$), we have $\D \kappa / \bar{\kappa} \simeq - 0.32 \, m / \bar{M}$.

Our derivation of the formulas \eqref{DA}, \eqref{DomegaH} and \eqref{Dkappa} for the surface area, angular frequency, and surface gravity of a black hole with moon relied on two variational relations comparing two nearby such spacetimes. Furthermore, we made extensive use of Stokes' theorem to relate horizon properties to quantities defined at infinity.  Alternatively, it would be interesting to recover these results by working directly with the metric of a single black-hole-moon spacetime, using explicit expressions (not currently known) for the behavior of the metric perturbation near the horizon, together with its asymptotic properties.

\section{Physical Process}\label{sec:physical}

The main result of this paper is the analytically computed, constant horizon surface gravity of a black hole with moon. In presenting this result as a \textit{change} in surface gravity due to the presence of a moon, we have adopted the conditions $\D M=e$ and $\D J=j$, enforcing that the metric perturbation contributes to the energy and angular momentum of the spacetime only through those of the moon, and not via any additional piece of the Kerr spacetime.\footnote{If instead one demands $\D M = e + e'$ and $\D J = j + j'$ for some $e'$ and $j'$, the formula for the perturbed surface gravity is modified by the addition of the surface gravity of a Kerr hole of mass $e'$ and angular momentum $j'$.} These conditions also arise from the demand that a locally defined Komar-type mass and angular momentum of the black hole are unmodified by the metric perturbation \cite{Ke.al2.10}. Finally, as we have shown, these conditions also entail that the surface area/entropy perturbation vanishes [recall Eqs.~\eqref{1st_law_bis} and \eqref{DA}]. For these reasons, our (standard) conditions naturally capture the idea of perturbing to the ``same'' black hole.

Nevertheless, the reader may wonder how, in principle, a physical process could realize the type of comparison between black-hole-with-moon and black-hole-without-moon that our conditions entail. One method of doing so begins with an operator, armed with moon, hovering at a great distance above the black hole, along its symmetry axis.  Using a flexible, adjustably tensile rod, the operator should extend the moon outwards, perpendicularly to the symmetry axis, and place it on a circular (nongeodesic) trajectory of angular frequency equal to that of the black hole horizon. While the moon (and rod) emit significant radiation during this process, only a negligible fraction will be absorbed by the distant black hole.

Once the moon is placed on this circular orbit, the operator should slowly lower the trajectory and adjust its radius until the corotating geodesic orbit is reached, while taking care to maintain the angular frequency equal to that of the black hole event horizon. There will be negligible radiation associated with the quasi-stationary lowering (and widening/narrowing) of the orbit, while there will be significant radiation associated with the circular motion. However, this radiation will have a helical symmetry adapted to the black hole, and no energy or angular momentum will flow across the horizon. Finally, the operator should release the moon onto the geodesic orbit and retract his rod. During the entire process of adding the moon, no energy or angular momentum was added to the black hole, and it is therefore the ``same'' black hole in the sense of our conditions. If the operator is equipped with a (quantum) particle detector and measures the black hole temperature before and after the lowering process, he should find our formula \eqref{Dkappa} for the difference. In this sense the moon does, in fact, cool the black hole.

\begin{acknowledgments}
It is a pleasure to thank John Friedman, Ted Jacobson, Eric Poisson, and Bob Wald for helpful discussions. S.G. acknowledges support from NASA through the Einstein Fellowship Program, Grant PF1-120082. A.L.T. acknowledges support from NSF through Grants PHY-0903631 and PHY-1208881, and from the Maryland Center for Fundamental Physics.
\end{acknowledgments}

\bibliography{}

\end{document}